\documentclass{ws-rv975x65}
\usepackage{ws-rv-van}
\makeindex

\def\etal{et al.}

\begin{document}

\chapter{A metal-insulator transition in 2D:\\ Established facts and open questions\label{ch1}}

\author{S.~V. Kravchenko}

\address{Physics Department, Northeastern University, Boston, MA 02115, USA\\
s.kravchenko@neu.edu}

\author[S.~V. Kravchenko and M.~P. Sarachik]{M.~P. Sarachik}

\address{Physics Department, City College of the City University of New York, New York, NY 10031, USA\\
sarachik@sci.ccny.cuny.edu}

\begin{abstract}
The discovery of a metallic state and a metal-insulator transition (MIT) in two-dimensional (2D) electron systems challenges one of the most influential paradigms of modern mesoscopic physics, namely, that ``there is no true metallic behavior in two dimensions''.  However, this conclusion was drawn for systems of noninteracting or weakly interacting carriers, while in all 2D systems exhibiting the metal-insulator transition, the interaction energy greatly exceeds all other energy scales. We review the main experimental findings and show that, although significant progress has been achieved in our understanding of the MIT in 2D, many open questions remain.
\end{abstract}
\vspace{1cm}

\body

In 2D electron systems, the electrons move in a plane in the presence of a weak random potential.  According to the scaling theory of localization of Abrahams \textit{et al}.\ \cite{abrahams79}, these systems lie on the boundary between high and low dimensions insofar as the metal-insulator transition is concerned.  The carriers are always strongly localized in one dimension, while in three dimensions, the electronic states can be either localized or extended.  In the case of two dimensions the electrons may conduct well at room temperature, but a weak logarithmic increase of the resistance is expected as the temperature is reduced.  This is due to the fact that, when scattered from impurities back to their starting point, electron waves interfere constructively with their time reversed paths.  Quantum interference becomes increasingly important as the temperature is reduced and leads to localization of the electrons, albeit on a large length scale; this is generally referred to as ``weak localization''. Indeed, thin metallic films and many of the two-dimensional electron systems fabricated on semiconductor surfaces display the predicted logarithmic increase of resistivity.

The scaling theory \cite{abrahams79} does not consider the effects of the Coulomb interaction between electrons.  The strength of the interactions is usually characterized by the dimensionless Wigner-Seitz radius, $$r_s=\frac{1}{(\pi n_s)^{1/2}a_B}$$ (here $n_s$ is the electron density and $a_B$ is the Bohr radius in a semiconductor).  As the density of electrons is reduced, the Wigner-Seitz radius grows and the interactions provide the dominant energy of the system.  In the early 1980's, Finkelstein \cite{finkelstein83,finkelstein84b} and Castellani \textit{et al}.\ \cite{castellani84} found that for weak disorder and sufficiently strong interactions, a 2D system scales towards a {\em conducting} state as the temperature is lowered.  However, the scaling procedure leads to an increase in the effective strength of the spin-related interactions and to a divergent spin susceptibility, so that the perturbative approach breaks down as the temperature is reduced toward zero.  Therefore, the possibility of a 2D metallic ground state stabilized by strong electron-electron interactions was not seriously considered at that time, particularly as there were no experimental observations to support the presence of a metallic phase.

Progress in semiconductor technology has enabled the fabrication of high quality 2D samples with very low randomness in which measurements can be made at very low carrier densities.  The strongly-interacting regime ($r_s\gg1$) has thus become experimentally accessible.  The first observation of a metal-insulator transition in strongly-interacting, low-disordered 2D systems on a silicon surface was reported in 1987 by Zavaritskaya and Zavaritskaya \cite{zavaritskaya87}.  Although identified by the authors as a metal-insulator transition, the discovery went by unnoticed.  Subsequent experiments on even higher mobility silicon samples \cite{kravchenko94,kravchenko95a,kravchenko96,popovic97,mcfarland09} confirmed the existence of a metal-insulator transition in 2D and demonstrated that there were surprising and dramatic differences between the behavior of strongly interacting systems with $r_s>10$ as compared with weakly-interacting systems. These results were met with great skepticism and were largely overlooked until they were confirmed in other strongly-interacting 2D systems in 1997 \cite{coleridge97,hanein98a,hanein98b,papadakis98,okamoto04,lai05}.  Moreover, it was found \cite{dolgopolov92,simonian97b,pudalov97,simmons98} that in the strongly-interacting regime, an external in-plane magnetic field strong enough to polarize the spins of the electrons or holes induces a giant positive magnetoresistance and completely suppresses the metallic behavior, implying that the spin state is central to the high conductance of the metallic state. Recent experiments \cite{okamoto99,kravchenko00b,shashkin01a,vitkalov01b,shashkin02,vitkalov02,pudalov02b,zhu03} have shown that there is a sharp enhancement of the spin susceptibility as the metal-insulator transition is approached. Interestingly, this enhancement is due to a strong increase of the effective mass, while the g-factor remains essentially constant \cite{shashkin02,pudalov02b,shashkin03,anissimova06}. Therefore, the effect is not related to the Stoner instability \cite{stoner46}.

In this article we summarize the main experimental findings. Of the many theories that have been proposed to explain the observations, we provide a detailed discussion of the theory of Punnoose and Finkelstein,\cite{punnoose05} as it provides numerical predictions with which experimental results can be compared directly. We end with a brief discussion of some of the unsolved problems.

\section{EXPERIMENTAL RESULTS IN ZERO MAGNETIC FIELD}

The first experiments that demonstrated the unusual temperature dependence of the resistivity \cite{zavaritskaya87,kravchenko94,kravchenko95a,kravchenko96,popovic97} were performed on low-disordered MOSFETs with maximum electron mobilities
reaching more than $4\times10^4$~cm$^2$/Vs; these mobilities were considerably higher than in samples used in earlier investigations. The very high quality of the samples allowed access to the physics at electron densities below $10^{11}$~cm$^{-2}$.  At these low densities, the Coulomb energy, $E_C$, is the dominant parameter. Estimates for Si MOSFETs at $n_s=10^{11}$~cm$^{-2}$ yield $E_C\approx10$~meV, while the Fermi energy, $E_F$, is about $0.6$~meV (a valley degeneracy of two is taken into account when calculating the Fermi energy, and the effective mass is assumed to be equal to the band mass.) The ratio between the Coulomb and Fermi energies, $r^*\equiv E_C/E_F$, thus assumes values above 10 in these samples.

\begin{figure}
\begin{center}
  \parbox{2in}{\epsfig{figure=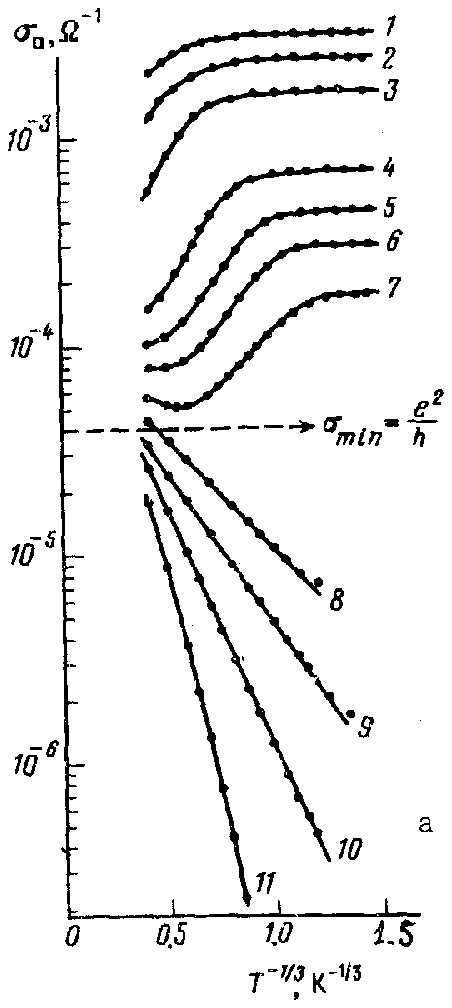,width=1.35in}
  \figsubcap{a}}
  \parbox{2in}{\epsfig{figure=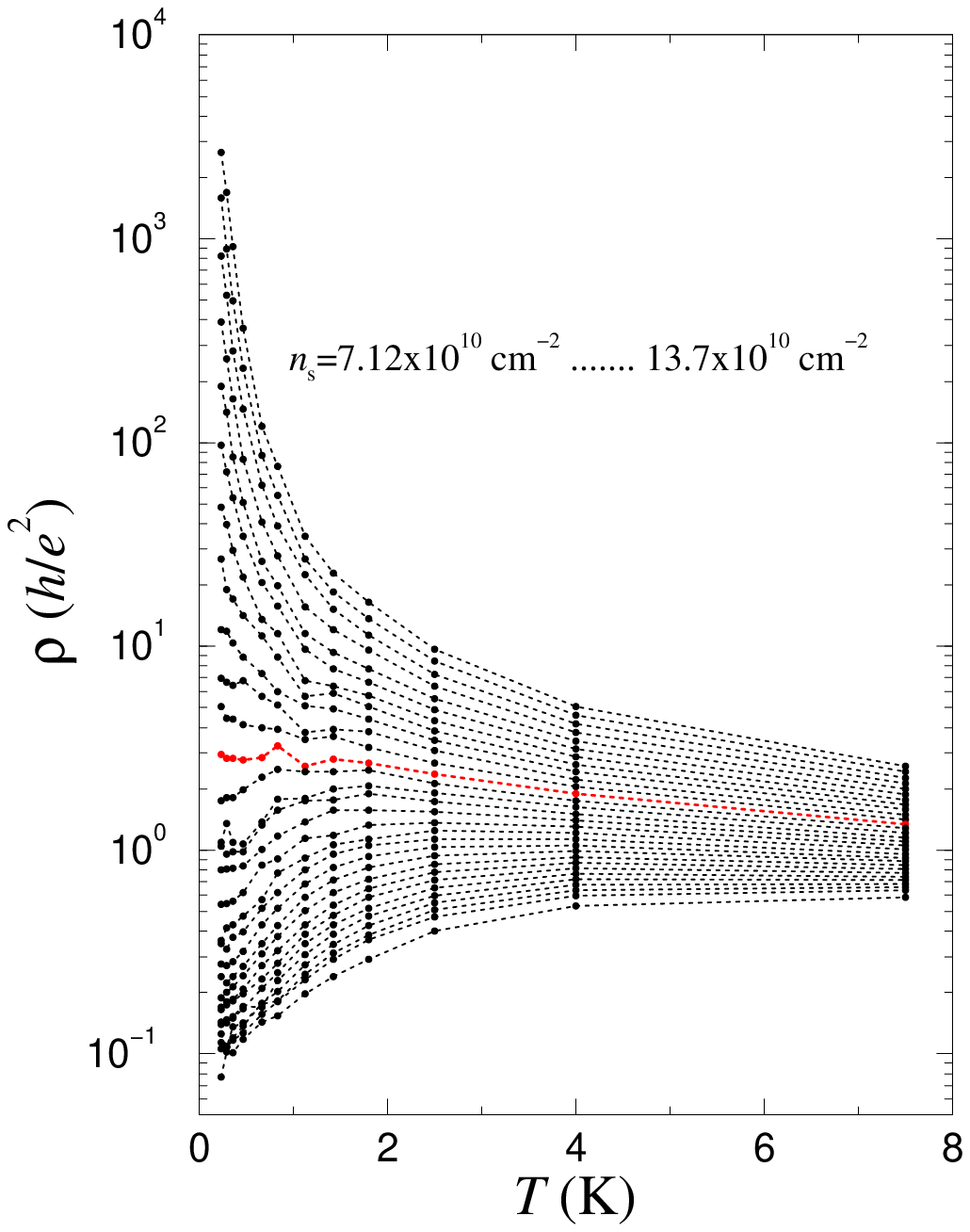,width=2.3in}
 \figsubcap{b}}
\end{center}
\caption{\label{r(t)first} (a) Conductivity \textit{vs}.\ the inverse cube root of temperature in silicon inversion channel for electron densities, $n_s$ ranging from $10^{12}$~cm$^{-2}$ (the upper curve) to less than $10^{11}$ ~cm$^{-2}$ (the lowest curve); adapted from Ref.\cite{zavaritskaya87}.  (b)~Temperature dependence of the $B=0$ resistivity in a dilute low-disordered Si MOSFET for 30 different electron densities ranging from $7.12\times10^{10}$~cm$^{-2}$ to $13.7\times10^{10}$~cm$^{-2}$; adapted from Ref.\cite{kravchenko95a}.}
\end{figure}

The earliest data that clearly show the MIT in 2D are shown in Fig.\ref{r(t)first}~(a).  Depending on the initial (``high-temperature'') value of conductivity, $\sigma_0$, the temperature dependence of conductivity $\sigma(T)$ in a low-disordered Si MOSFET exhibits two different types of behavior: for $\sigma_0<e^2/h$, the conductivity decreases with decreasing temperature following Mott's hopping law in 2D, $\sigma\propto\exp T^{-1/3}$; on the other hand, for  $\sigma_0>e^2/h$, the conductivity increases with decreasing $T$ by as much as a factor of 7 before finally saturating at sub-kelvin temperatures. Fig.~\ref{r(t)first}~(b) shows the temperature dependence of the resistivity (the inverse of the conductivity) measured in units of $h/e^2$ of a high-mobility MOSFET for 30 different electron densities $n_s$ varying from $7.12\times10^{10}$ to $13.7\times10^{10}$~cm$^{-2}$.  If the resistivity at high temperatures exceeds the quantum resistance $h/e^2$, $\rho(T)$ increases monotonically as the temperature decreases, behavior that is characteristic of an insulator.  However, for $n_s$ above a certain ``critical'' value, $n_c$ (the curves below the ``critical" curve that extrapolates to $3h/e^2$), the temperature dependence of $\rho(T)$ is non-monotonic: with decreasing temperature, the resistivity first increases (at $T>2$~K) and then decreases as the temperature is further reduced.  At yet higher density $n_s$, the resistivity is almost constant at $T>4$~K but drops by an order of magnitude at lower temperatures, showing strongly metallic behavior as $T\rightarrow0$.

\begin{figure}\hspace{-1mm}
\begin{center}
\scalebox{.37}{\includegraphics{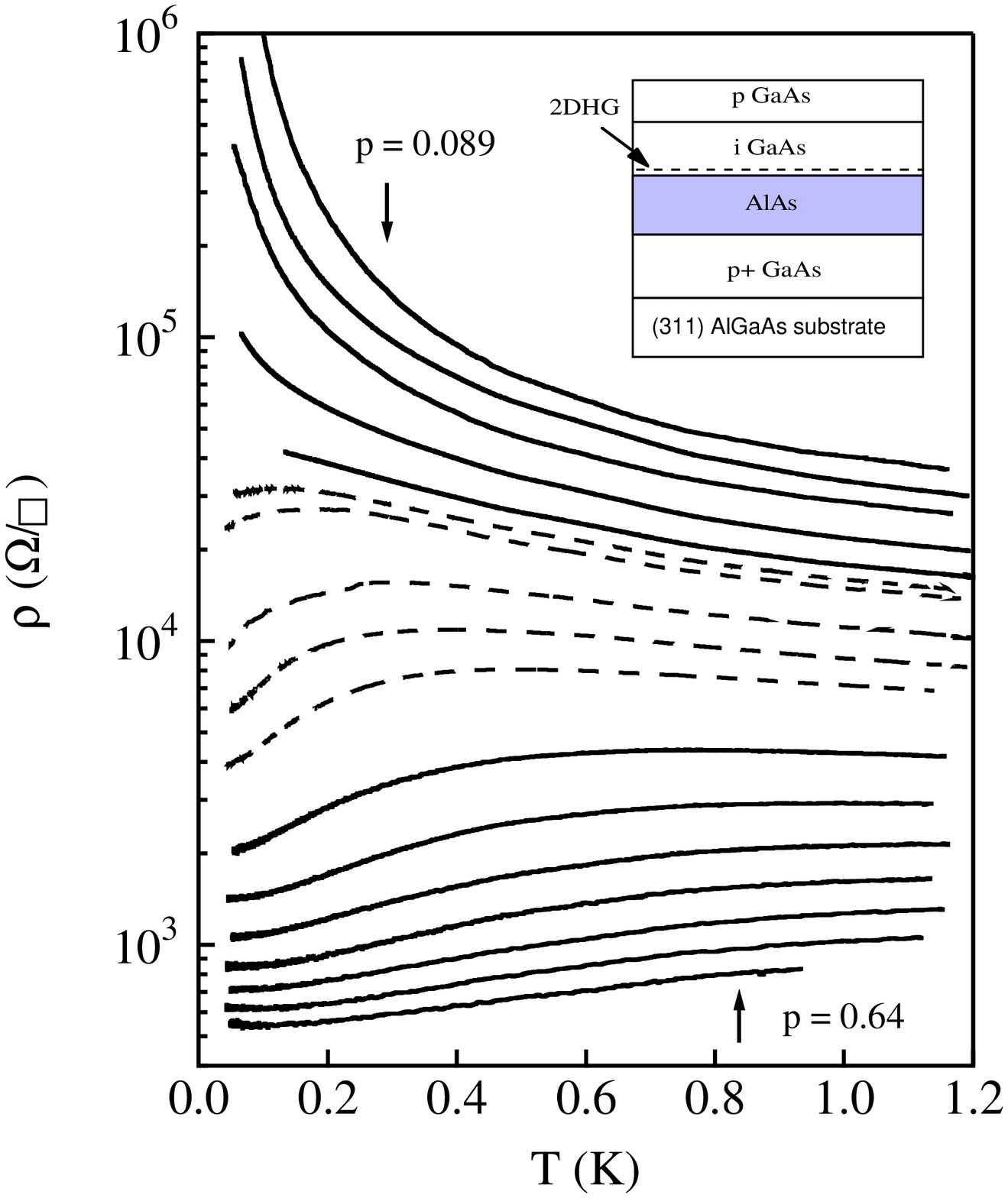}}
\end{center}
\caption{\label{r(t)gaas} For low-disordered 2D hole systems in p-GaAs/AlGaAs, the resistivity per square is shown as a function of temperature for $B=0$ at various fixed hole densities, $p$.  Data for an ISIS (inverted semiconductor-insulator-semiconductor) structure with hole densities (from top to bottom) $p=$~0.89, 0.94, 0.99, 1.09, 1.19, 1.25, 1.30, 1.50, 1.70, 1.90, 2.50, 3.20, 3.80, 4.50, 5.10, 5.70, and $6.40\cdot10^{10}$~cm$^{-2}$.  The inset shows a schematic diagram of the ISIS structure: The carriers are accumulated in an undoped GaAs layer situated on top of an undoped AlAs barrier, grown over a $p^+$ conducting layer which serves as a back-gate; the hole density, $p$, is varied by applying a voltage to the back gate.  From Ref.\cite{hanein98a}.}
\end{figure}

A metal-insulator transition similar to that seen in clean silicon MOSFETs has also been observed in other low-disordered, strongly-interacting 2D systems: $p$-type SiGe heterostructures \cite{coleridge97,senz99}, p-GaAs/AlGaAs heterostructures \cite{hanein98a,yoon99,mills99}, n-GaAs/AlGaAs heterostructures \cite{hanein98b,lilly03}, AlAs heterostructures \cite{papadakis98}, and n-SiGe heterostructures \cite{okamoto04,lai05}.  The values of the resistivity are quite similar in all systems.  In Fig.~\ref{r(t)gaas}, the resistivity is shown as a function of temperature for a $p$-type GaAs/AlGaAs sample; here the interaction parameter, $r_s$, changes between approximately 12 and 32.\footnote{These $r_s$ values were calculated assuming that the effective mass is independent of density and equal to $0.37\,m_e$, where $m_e$ is the free-electron mass.}  The main features are very similar to those found in silicon MOSFETs: when the resistivity at ``high'' temperatures exceeds the quantum resistance, $h/e^2$ ({\it i.e.}, at hole densities below some critical value, $p_c$), the $\rho(T)$ curves are insulating-like in the entire temperature range; for densities just above $p_c$, the resistivity shows insulating-like behavior at higher temperatures and then drops by a factor of 2 to 3 at temperatures below a few hundred mK; and at yet higher hole densities, the resistivity is metallic in the entire temperature range.  Note that the curves that separate metallic and insulating behavior have resistivities that increase with decreasing temperature at the higher temperatures shown; this is quite similar to the behavior of the separatrix in silicon MOSFETs when viewed over a broad temperature range (see Fig.~\ref{r(t)first}).  Below approximately 150~mK, the separatrix in $p$-type GaAs/AlGaAs heterostructures is independent of temperature \cite{hanein98a}, as it is in Si MOSFETs below approximately 2~K.  The resistivity of the separatrix in both systems extrapolates to $\approx 2$ or $3\,h/e^2$ as $T\rightarrow0$, even though the corresponding carrier densities are very different.

\section{THE EFFECT OF A MAGNETIC FIELD}

\begin{figure}
\scalebox{.4}{\includegraphics{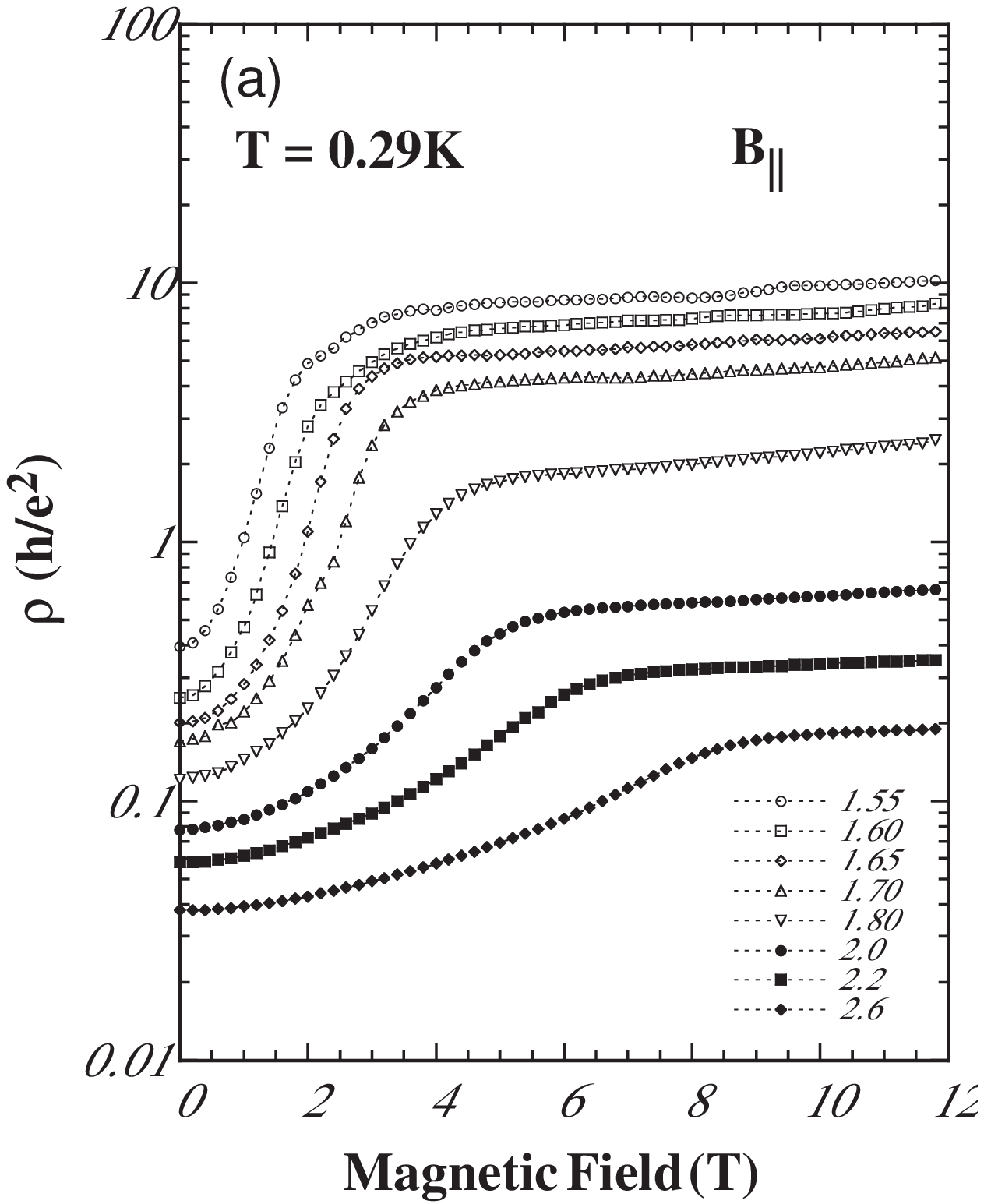}}
\scalebox{.35}{\includegraphics{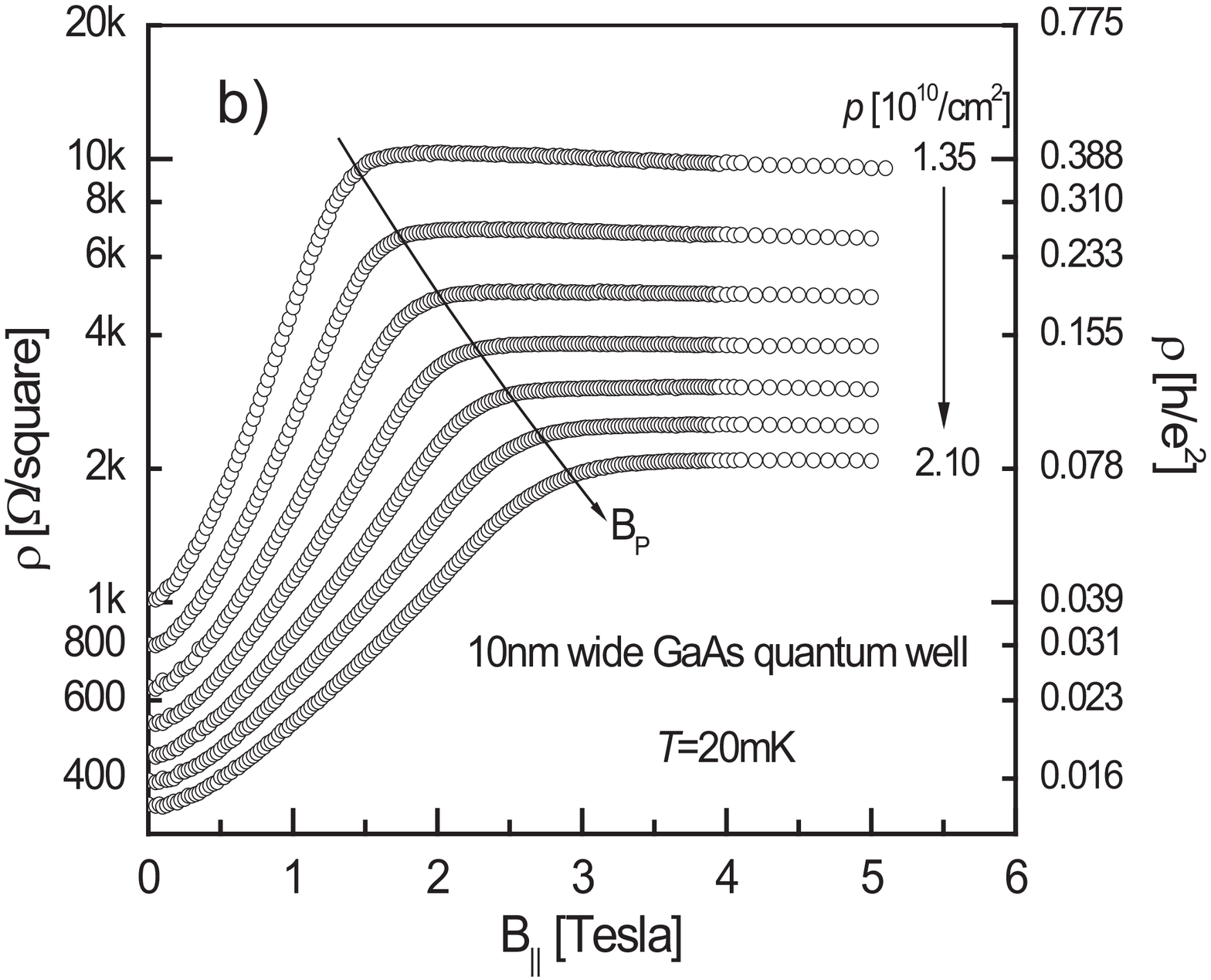}}
\caption{\label{RB_pudalov} (a) Resistivity versus parallel magnetic field measured at $T=0.29$~K in a Si MOSFET.  Different symbols correspond to densities from 1.01 to $2.17\cdot10^{11}$~cm$^{-2}$; adapted from Ref.\cite{pudalov97}  (b) Resistivity as a function of $B_\parallel$ of a 10 nm wide p-GaAs quantum well at 50~mK; adapted from Ref.\cite{gao06}.}
\end{figure}

In ordinary metals, the application of a parallel magnetic field ($B_\parallel$) does not lead to any dramatic changes in the transport properties: if the thickness of the 2D electron system is small compared to the magnetic length, the parallel field couples largely to the electrons' spins while the orbital effects are suppressed.  Only weak corrections to the conductivity are expected due to electron-electron interactions \cite{lee85}.  It therefore came as  a surprise when Dolgopolov~\textit{et al.} \cite{dolgopolov92} observed a dramatic suppression of the conductivity in dilute Si MOSFETs by a parallel in-plane magnetic field $B_\parallel$.  The magnetoresistance in a parallel field was studied in detail by Simonian~\textit{et al.}  \cite{simonian97b} and Pudalov~\textit{et al.} \cite{pudalov97}, also in Si MOSFETs.  In the left hand part of Fig.~\ref{RB_pudalov}, the resistivity is shown as a function of parallel magnetic field at a fixed temperature of 0.3~K for several electron densities.  The resistivity increases sharply as the magnetic field is raised, changing by a factor of about 4 at the highest density shown and by more than an order of magnitude at the lowest density, and then saturates and remains approximately constant up to the highest measuring field, $B_\parallel=12$~tesla.  The magnetic field where the saturation occurs, $B_{\rm sat}$, depends on $n_s$, varying from about 2~tesla at the lowest measured density to about 9 tesla at the highest.  The metallic conductivity is suppressed in a similar way by magnetic fields applied at any angle relative to the 2D plane \cite{kravchenko98} independently of the relative directions of the measuring current and magnetic field \cite{simonian97b,pudalov02a}.  All these observations suggest that the giant magnetoresistance is due to coupling of the magnetic field to the electrons' spins.  Indeed, from an analysis of the positions of Shubnikov-de~Haas oscillations in tilted magnetic fields \cite{okamoto99,vitkalov00,vitkalov01a} it was concluded that in MOSFETs, the magnetic field $B_{\rm sat}$ is equal to that required to fully polarize the electrons' spins.

In $p$-type GaAs/AlGaAs heterostructures, the effect of a parallel magnetic field is qualitatively similar, as shown in the right hand part of Fig.~\ref{RB_pudalov}.  As in the case of Si MOSFETs, there is a distinct knee that serves as a demarcation between the behavior in low and high fields.  For high hole densities, Shubnikov-de~Haas measurements \cite{tutuc01} have shown that this knee is associated with full polarization of the spins by the in-plane magnetic field.  However, unlike Si MOSFETs, the magnetoresistance in p-GaAs/AlGaAs heterostructures has been found to depend on the relative directions of the measuring current, magnetic field, and crystal orientation \cite{papadakis00}; one should note that the crystal anisotropy of this material introduces added complications.  In p-SiGe heterostructures, the parallel field was found to induce negligible magnetoresistance \cite{senz99} because in this system the parallel field cannot couple to the spins due to very strong spin-orbit interactions.

\begin{figure}[h]
\begin{center}
  \scalebox{.7}{\epsfig{figure=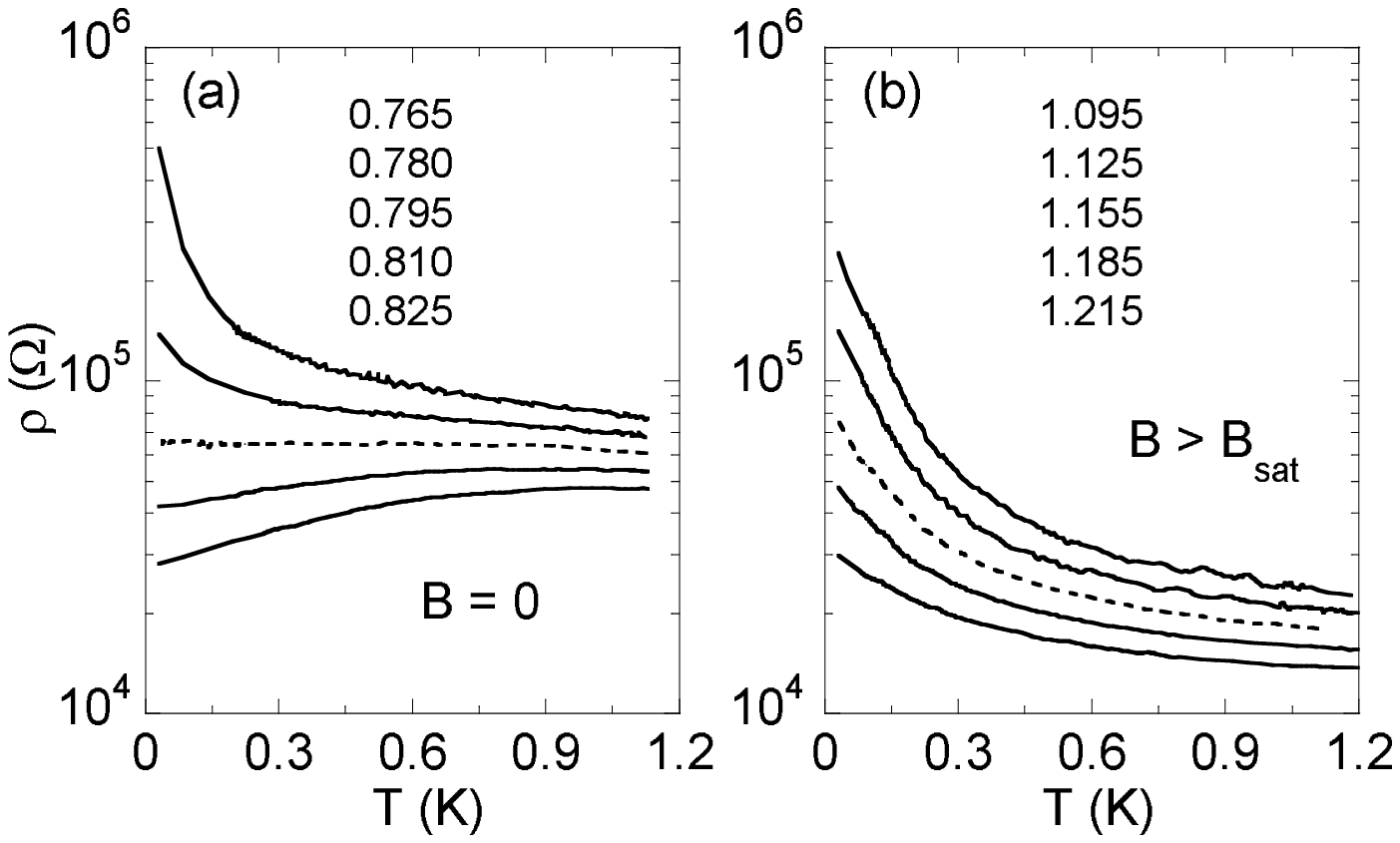}}\\
  \vspace{.2in}
  \parbox{1.8in}{\epsfig{figure=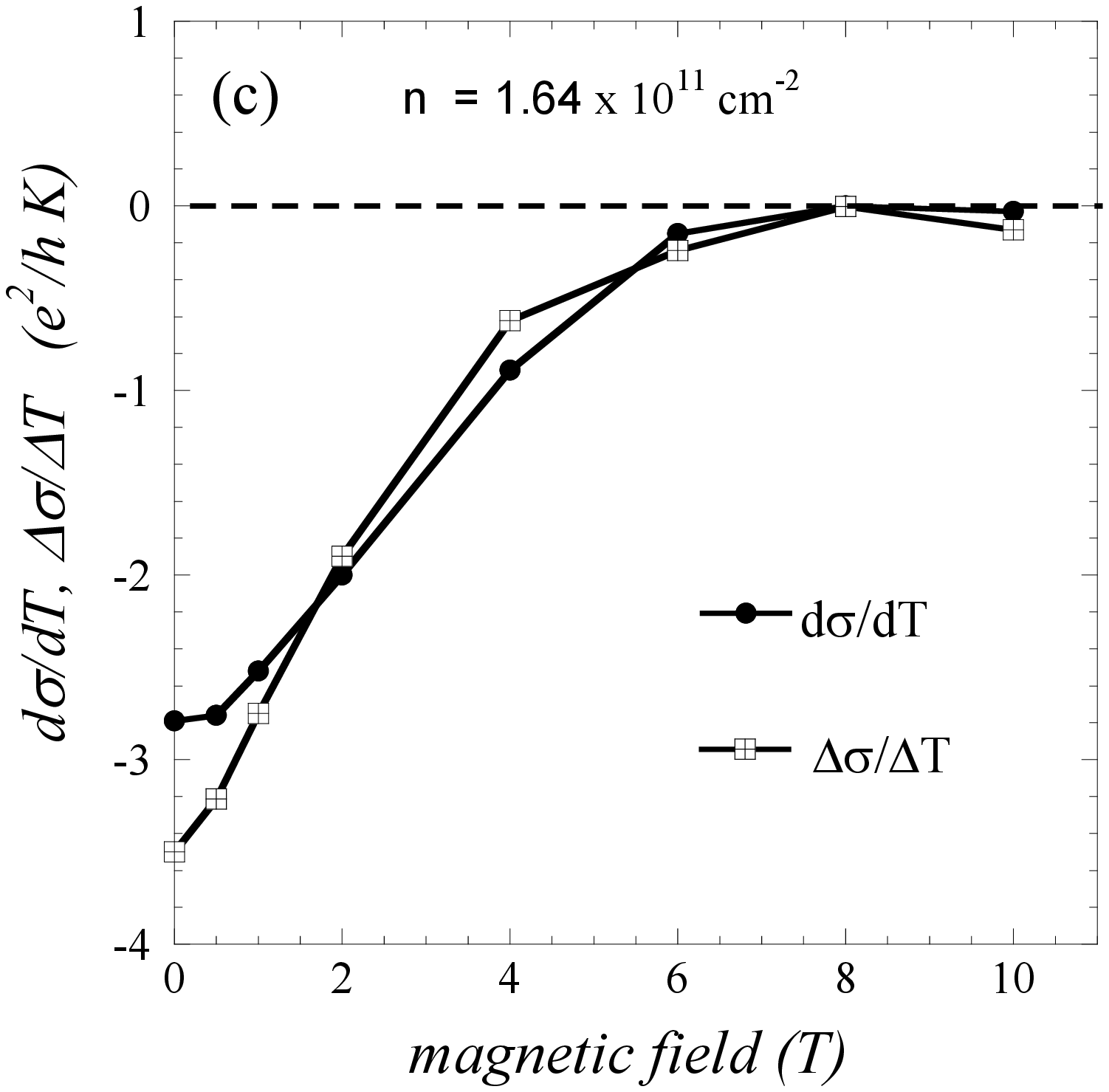,width=2.3in}}
\caption{Temperature dependence of the resistivity of a Si MOSFET at different electron densities near the MIT in zero magnetic field (a), and in a parallel magnetic field of 4~tesla (b). The electron densities are indicated in units of $10^{11}$~cm$^{-2}$. Dashed curves correspond to $n_s=n_{c1}$ which is equal to $0.795\times10^{11}$~cm$^{-2}$ in zero field and to $1.155\times10^{11}$~cm$^{-2}$ in $B_\parallel=4$~tesla; taken from Ref.\cite{shashkin01b}.   (c) Data taken as a function of parallel magnetic field for a silicon MOSFET of density $1.64 \times 10^{11}$ cm$^{-2}$; closed symbols denote the slope $d\sigma/dT$ and open symbols denote $\Delta \sigma/\Delta T$ calculated for the temperature interval $0.27$ to $1.35$ K; from Ref.\cite{tsui05}}
\label{R(T)inB_ours}
\end{center}
\end{figure}

Over and above the very large magnetoresistance induced by an in-plane magnetic fields, an even more important effect of a parallel field is that it causes the zero-field 2D metal to become an insulator \cite{simonian97b,mertes01,shashkin01b,gao02,tsui05}.  The extreme sensitivity to parallel field is illustrated in Fig.~\ref{R(T)inB_ours}. The top two panels compare the temperature dependence of the resistivity in the absence and in the presence of a parallel magnetic field.  For $B_\parallel=0$, the resistivity displays the familiar, nearly symmetric (at temperatures above 0.2~K) critical behavior about the separatrix (the dashed line).  However, in a parallel magnetic field of $B_\parallel=4$~tesla, which is high enough to cause full spin polarization at this electron density, all the $\rho(T)$ curves display ``insulating-like'' behavior, including those which start below $h/e^2$ at high temperatures.  There is no temperature-independent separatrix at any electron density in a spin-polarized electron system \cite{simonian97b,shashkin01b}.  The effect of a parallel magnetic field is further demonstrated in the bottom panel of Fig.~\ref{R(T)inB_ours}, where the slope of the resistivity calculated for the temperature interval $0.27$ K to $1.35$ K is plotted as a function of magnetic field at a fixed density; these data show explicitly and quantitatively that a magnetic field applied parallel to the plane of the electrons reduces the temperature dependence of the conductivity to near zero.  Moreover, this was shown to be true  over a broad range of electron densities extending deep into the metallic regime where the high-field conductivity is on the order of $10 (e^2/h)$. The clear difference in behavior with and without in-plane magnetic field convincingly demonstrates that the spin-polarized and unpolarized states behave very differently and rules out explanations that predict similar behavior of the resistance regardless of the degree of spin polarization.

\section{SPIN SUSCEPTIBILITY NEAR THE METAL-INSULATOR TRANSITION}
\subsection{Experimental measurements of the spin susceptibility}\label{spin}

In Fermi-liquid theory, the electron effective mass and the $g$-factor (and, therefore, the spin susceptibility, $\chi\propto g^*m^*$) are renormalized due to electron-electron interactions \cite{landau57}. Earlier experiments \cite{fang68,smith72}, performed at relatively small values of $r_s \sim 2$ to 5, confirmed the expected increase of the spin susceptibility. More recently, Okamoto~\etal \cite{okamoto99} observed a renormalization of $\chi$ by a factor of $\sim2.5$ at $r_s$ up to about 6.  At yet lower electron densities, in the vicinity of the metal-insulator transition, Kravchenko~\etal \cite{kravchenko00b} have observed a disappearance of the energy gaps at ``cyclotron'' filling factors which they interpreted as evidence for an increase of the spin susceptibility by a factor of at least five.

\begin{figure}[h]
\begin{center}
\hspace*{-.5in}
  \parbox{2.45in}{\epsfig{figure=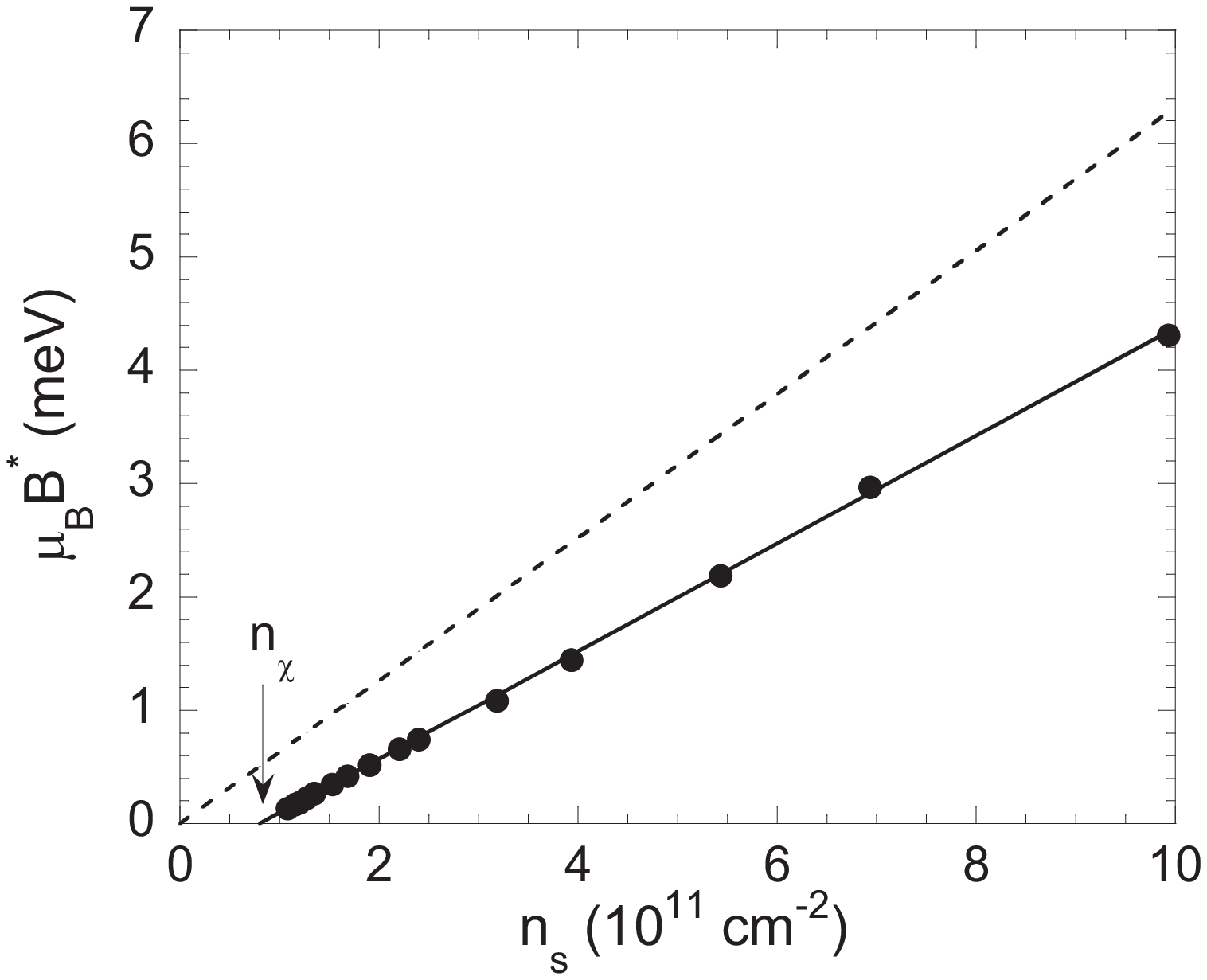,width=2.45in}
  \figsubcap{a}}
  \parbox{2in}{\epsfig{figure=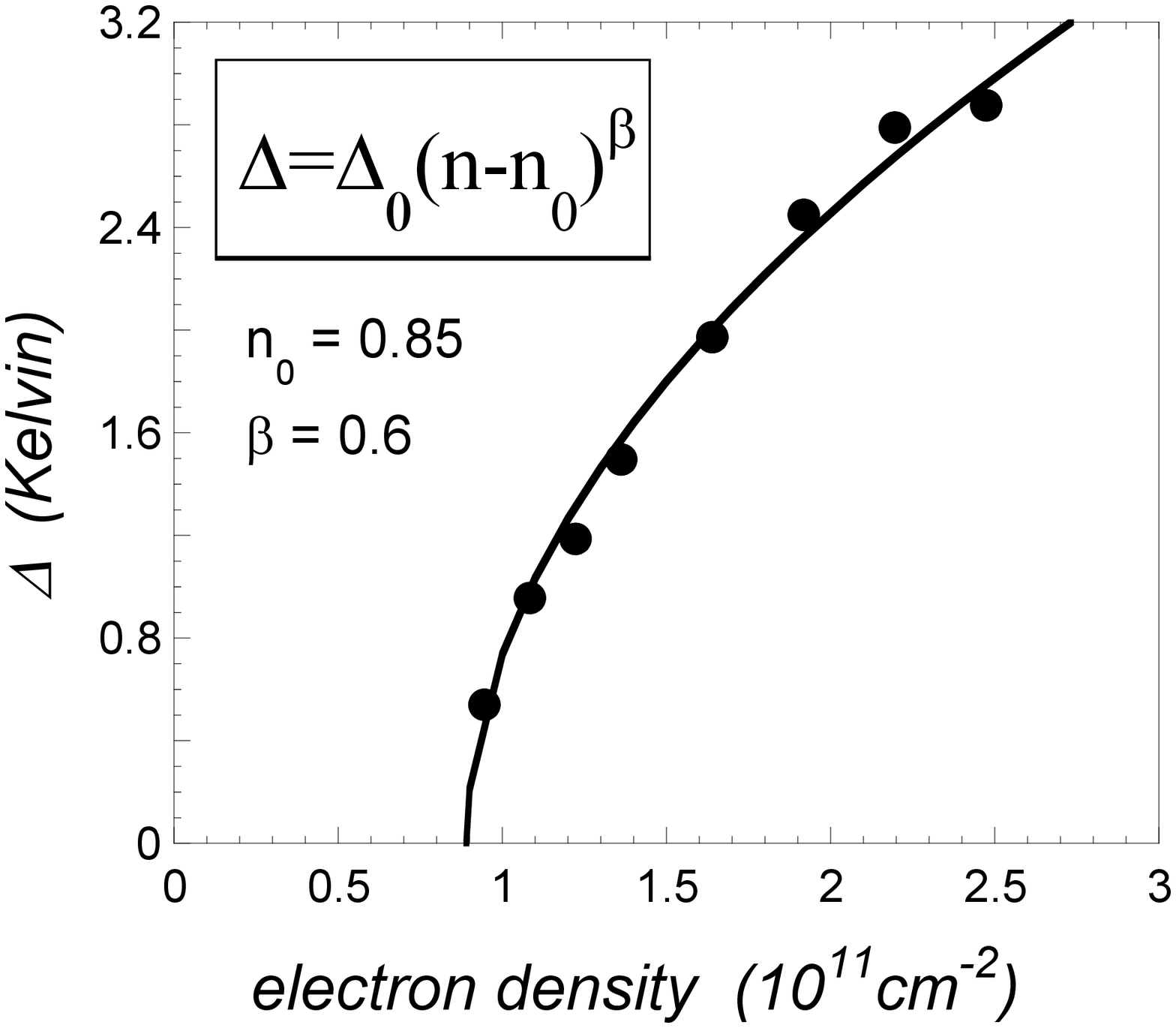,width=2.3in}
 \figsubcap{b}}
\caption{(a) Magnetic field for the onset of complete spin polarization \textit{vs}.\ electron density; the dashed line depicts $B^\ast$ calculated assuming that $g$ and $m$ are not renormalized; adapted from Ref.\cite{shashkin01a}. (b) Characteristic energy $k_B\Delta$ associated with the magnetic field dependence of the conductivity plotted as a function of electron density; the parameter  $\Delta$ decreases
with decreasing density, and extrapolates to 0 at a critical density $n_o$; adapted from Ref.\cite{vitkalov01b}.}
\label{fig:Bstar}
\end{center}
\end{figure}

It was noted many years ago by Stoner that strong interactions can drive an electron system toward a ferromagnetic instability \cite{stoner46}. Within some theories of strongly interacting 2D systems \cite{finkelstein83,finkelstein84a,finkelstein84b,castellani84,punnoose05}, a tendency toward ferromagnetism is expected to accompany metallic behavior. The easiest way to estimate the spin magnetization of 2D electrons (or holes) is to measure the magnetic field above which the magnetoresistance saturates (and thus full spin polarization is reached) as a function of electron density. For non-interacting electrons, the saturation field is proportional to the electron density:
	\[B^\ast=\frac{\pi\hbar^2n_s}{gm\mu_B}.
\]
Here $g$ is the Land\'e $g$-factor, $m$ is the effective mass, and $\mu_B$ is the Bohr magneton. Experiments \cite{shashkin01a,vitkalov01b} have shown, however, that in strongly correlated 2D systems in Si MOSFETs, the parallel field required for full spin polarization extrapolates to zero at a non-zero electron density, $n_\chi$.  The left-hand panel of Fig.\ref{fig:Bstar} shows that the field $B^\ast$ for full polarization obtained by Shashkin \textit{et al.}\cite{shashkin01a} extrapolates to zero at a finite electron density; the dashed line indicates the calculated $B^*(n_s)$ for comparison; the fact that the measured $B^\ast$ lies significantly lower than the calculated value indicates that either $g$ or $m$ (or both) are larger than their band values.  Using a different method of analysis, Vitkalov \textit{et al.}\cite{vitkalov01b} obtained a characteristic energy $k_B\Delta$ associated with the magnetic field dependence of the conductivity plotted as a function of electron density, as shown in the right-hand panel of Fig.\ref{fig:Bstar}; the parameter  $\Delta$ decreases with decreasing density, and extrapolates to zero at a critical density labeled $n_o$.   That $B^\ast$ and $k_B\Delta$, both measures of the field required to obtain complete spin polarization, extrapolate to zero at a finite density implies  there is a spontaneous spin polarization at $n_s=n_\chi=n_0$. Many Si MOSFET samples of different quality have been tested, and the results indicate that $n_\chi\approx8\times10^{10}$~cm$^2$ is independent of disorder.  In the highest quality samples, $n_\chi$ was found to be within a few percent of the critical density for the metal-insulator transition, $n_c$ (but consistently below).

It is easy to recalculate the renormalized spin susceptibility using the data for $B^*(n_s)$: 
	\[\frac{\chi}{\chi_0}=\frac{n_s}{n_s-n_\chi},
\]
where $\chi_0$ is the ``non-interacting'' value of the spin susceptibility. Such critical behavior of a thermodynamic parameter usually indicates that a system is approaching a phase transition, possibly of magnetic origin. However, direct evidence of a phase transition can only be obtained from measurements of thermodynamic properties. Given the tiny number of electrons in a dilute 2D layer, magnetic measurements are very hard to perform. A clever technique was designed and implemented by Prus  \textit{et al.}\cite{prus02} and Shashkin \textit{et al.} \cite{shashkin06}. These authors modulated the parallel magnetic field with a small ac field, $B_{\text{mod}}$,  and measured the tiny induced current between the gate and the two-dimensional electron system. The imaginary (out-of-phase) component of the current is proportional to $d\mu/dB$, where $\mu$ is the chemical potential of the 2D gas. By applying the Maxwell relation $dM/dn_s=-d\mu/dB$, one can obtain the magnetization $M$ from the measured current. Full spin polarization corresponds to $dM/dn_s=0$. Yet another way of finding the density for complete spin polarization is related to measurements of the thermodynamic density of states of the 2D system obtained from measurements\cite{shashkin06} of the capacitance of a MOSFET: the thermodynamic density of states was found to change abruptly with the onset of complete spin polarization of the electrons' spins.

\begin{figure}
    \centering
         \includegraphics[width=.5\textwidth]{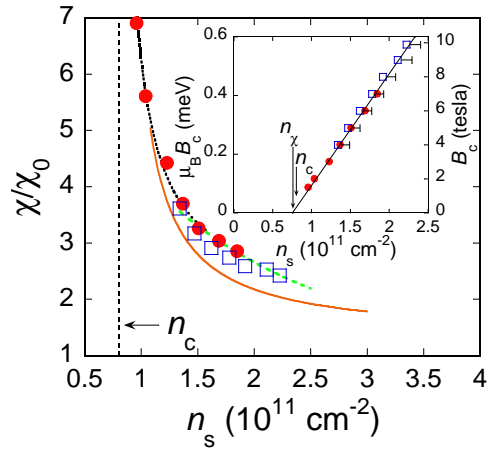}
\caption{The Pauli spin susceptibility as a function of electron density obtained by thermodynamic methods: direct measurements of the spin magnetization (dashed line), $d\mu/dB=0$ (circles), and density of states (squares). The dotted line is a guide to the eye. Also shown by a solid line is the transport data of Ref.\cite{shashkin01}. Inset: Field for full spin polarization as a function of the electron density determined from measurements of the magnetization (circles) and magnetocapacitance (squares). The data for $B_c$ are consistent with a linear fit which extrapolates to a density $n_\chi$ close to the critical density $n_c$ for the $B=0$ MIT. Adapted from Ref.\cite{shashkin06}.}
    \label{fig:spin}
\end{figure}

The results obtained for the spin susceptibility are shown in Fig.\ref{fig:spin}. One can see that upon approaching to the critical density of the metal-insulator transition, the spin susceptibility increases by almost an order of magnitude relative to its ``non-interacting'' value. This implies the occurrence of a spontaneous spin polarization (either Wigner crystal or ferromagnetic liquid) at low $n_s$, although in currently available samples, the formation of a band tail of localized electrons at $n_s\lesssim n_c$ conceals the origin of the low-density phase. In other words, so far, one can only reach an incipient transition to a new phase.

A strong dependence of the magnetization on $n$ has also been seen \cite{zhu03,vakili04,lu08} in other types of devices for $n_s$ near the critical density for the metal-insulator transition.

\subsection{Effective mass or $g$-factor?}

\begin{figure}
    \centering
         \includegraphics[width=.4\textwidth]{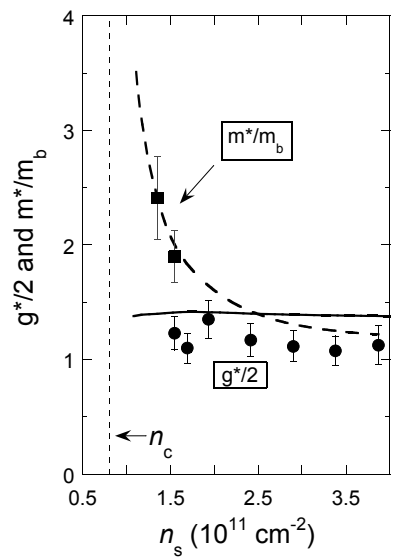}
\caption{The Land\'{e} $g$-factor and the effective mass as a function of electron density obtained from transport measurements \cite{shashkin02} (solid and dashed lines, respectively). Also shown are the effective $g$-factor (circles) and the cyclotron mass (squares) obtained by measurements of thermodynamic magnetization \cite{anissimova06}. The critical density $n_c$ for the metal-insulator transition is indicated by the arrow.}
    \label{fig:mass}
\end{figure}

In principle, the strong increase of the Pauli spin susceptibility at low electron densities can be due to either the increase of the effective mass or the Lande $g$-factor (or both). The effective mass was measured by several groups employing different methods \cite{shashkin02,shashkin03,anissimova06} which gave quantitatively similar results. The values $g/g_0$ and $m/m_b$ as a function of the electron density are shown in Fig. \ref{fig:mass} (here $g_0=2$ is the $g$ factor in bulk silicon, $m_b$ is the band mass equal to $0.19m_e$, and $m_e$ is the free electron mass).  In the high $n_s$ region (relatively weak interactions), the enhancement of both $g$ and $m$ is relatively small, with both values increasing slightly with decreasing electron density, in agreement with earlier data \cite{ando82}. Also, the renormalization of the $g$ factor is dominant compared to that of the effective mass, consistent with theoretical studies \cite{iwamoto91,kwon94,chen99}.

In contrast, the renormalization at low $n_s$ (near the critical region), where $r_s\gg1$, is striking. As the electron density is decreased, the effective mass increases dramatically while the $g$ factor remains essentially constant and relatively small, $g\approx g_0$. Hence, it is the effective mass, rather than the $g$ factor, that is responsible for the drastically enhanced spin susceptibility near the metal-insulator transition.

\subsection{Effective mass as a function of $r_s$}

\begin{figure}
\centering
\scalebox{0.3}{\includegraphics{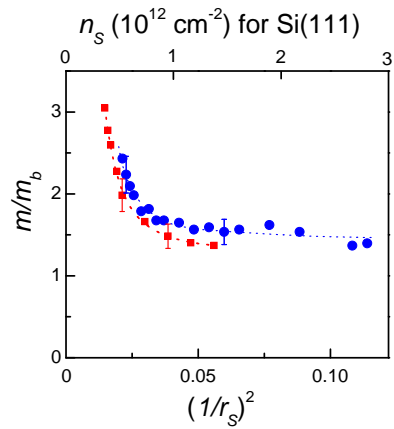}}
\caption{\label{fig:kwon} The effective mass (dots) in units of $m_b$ as a function of $(1/r_s)^2\propto n_s$. Also shown by squares is the data obtained in (100)-Si MOSFETs \cite{shashkin03}. The dashed lines are guides to the eye; from Ref.\cite{shashkin07}.}
\end{figure}

The effective mass has also been measured in a dilute 2D electron system in (111)-silicon. This system is interesting because the band electron mass $m_b=0.358m_e$ is  approximately a factor of two larger than it is in (100)-silicon. In addition, the (111)-silicon samples used in these experiments have a much higher level of disorder. Remarkably, the relative enhancement of the effective mass, \textit{i.e.}, $m^\ast/m_b$, was found to be essentially the same function of the interaction parameter, $r_s$, as in (100) samples of Si MOSFETs. Shown in Fig.~\ref{fig:kwon}, the effective mass plotted in units $m_b$ as a function of $(1/r_s)^2\propto n_s$ is essentially the same for the two systems within the experimental  uncertainty, despite the fact that the band mass differs by about a factor of two and the level disorder differs by almost one order of magnitude.  This implies that the relative mass
enhancement is determined solely by the strength of the electron-electron interactions.

\section{Comparison with theory and open questions}

Many theories have been proposed to account for the experimental observations summarized above.  These include exotic superconductivity,\cite{phillips98} the formation of a disordered Wigner solid,\cite{chakravarty99} microemulsion phases,\cite{spivak03,spivak04,spivak06} percolation,\cite{meir99} and a non-Fermi liquid state \cite{chakravarty98}.  In what follows, we restrict our discussion to theories that provide numerical predictions with which experimental can be compared directly.

\subsection{Ballistic regime ($k_B T\gg\hbar/\tau$)}

When the resistivity of a sample is much smaller than $h/e^2$, the electrons are in a ballistic regime for temperatures $T>h/k_B\tau$; this encompasses most of the experimentally accessible range except in samples with electron densities that are very close to the critical density.  Theories \cite{stern80,gold86,dassarma99,dolgopolov00,dassarma00,dassarma03,dassarma04} that invoke the effect of electron screening attempt to explain the transport results in the ballistic regime by extrapolating classical formulas for the resistivity, valid for $r_s<1$, to the regime where $r_{\text s} \gg 1$. Indeed, quantitatively successful comparisons with experiment have been reported.\cite{dassarma99,dassarma00,dassarma03,dassarma04} However, at large $r_{\text s}$, the screening length $\lambda_{\text {sc}}$ obtained using a random-phase approximation becomes parametrically smaller than the spacing between electrons: $\lambda_{\text {sc}}\sqrt{\pi n} = (1/4) r_{\text s}^{-1}  \ll 1$ \cite{spivak06}. Screening lengths smaller than the distance between electrons are clearly unphysical. 

This approach was corrected by Zala \textit {et al.}\cite{zala01}, who considered the contribution due to the scattering from the Friedel oscillations induced by impurities. At small $r_{\text s}$, the ``insulating-like'' sign of $d\rho/dT$ is obtained, in agreement with experiments on samples with $r_{\text s} \sim 1$.  However, when extrapolated to large enough $r_{\text s}$, $d\rho/dT$ changes sign, and $\rho(T)$ becomes a linearly increasing function of $T$. This theory \cite{zala01} predicts the complete suppression of the metallic behavior in parallel magnetic fields sufficiently strong to completely polarize spins, again in agreement with the experiments. However, one should keep in mind that this theory considers {\it corrections} to the conductivity that are small compared to the Drude conductivity. By contrast, changes in resistivity by an order of magnitude are often observed experimentally.

\subsection{Scaling theory of the metal-insulator transition in 2D: diffusive regime ($k_B T\ll\hbar/\tau$)}

\begin{figure*}
\centering \includegraphics[width=.4\textwidth]{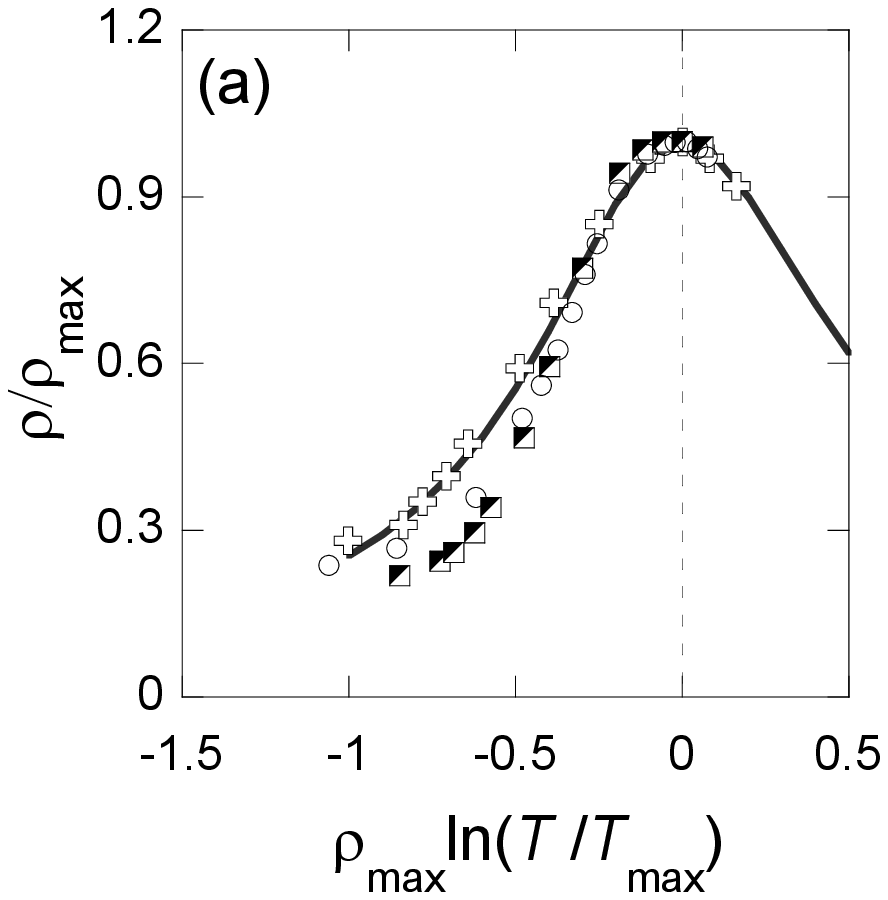}
 \includegraphics[width=.4\textwidth]{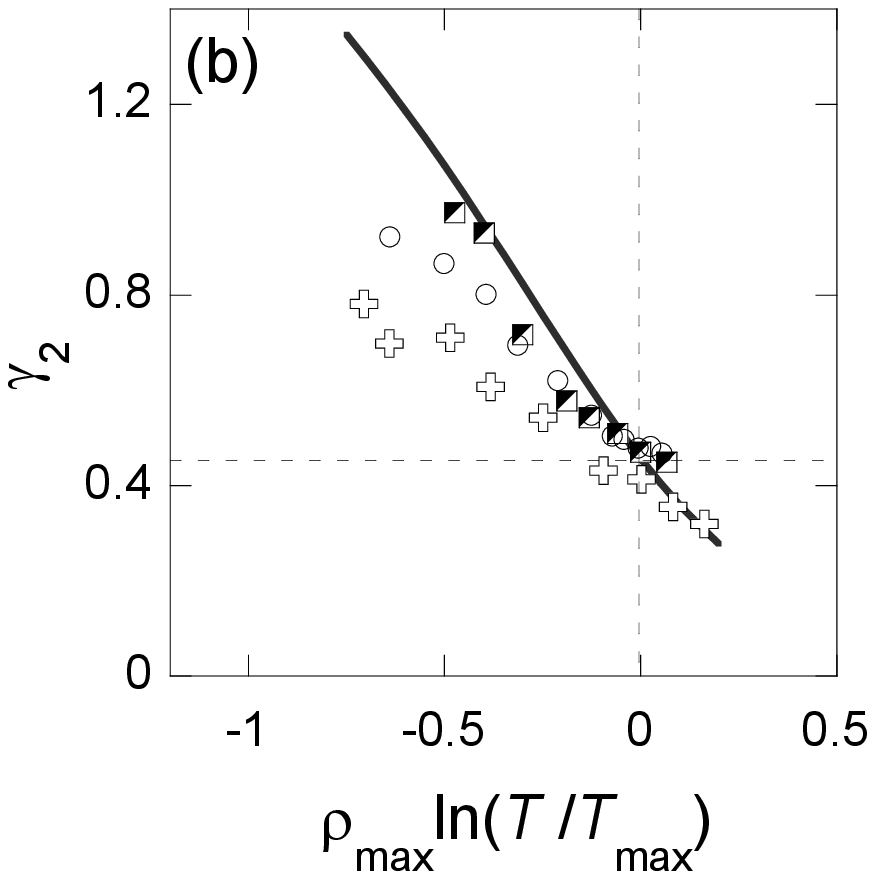}\\
\caption{Comparison between theory (lines) and experiment (symbols).  (a): $\rho/\rho_{\text{max}}$ as a function of $\rho_{\text{max}}\;\text{ln}(T/T_{\text{max}})$.  (b): $\gamma _{2}$ as a function of $\rho_{\text{max}}\;\text{ln}(T/T_{\text{max}})$. Vertical dashed lines correspond to $T=T_{\text{max}}$, the temperature at which $\rho(T)$ reaches  maximum.  Note that at this temperature, the interaction amplitude $\gamma_2\approx 0.45$ (indicated by the horizontal dashed line in (b)), in excellent agreement with theory. Electron densities are $9.87$ (squares), $9.58$ (circles), and $9.14\times 10^{10}$~cm$^{-2}$ (crosses). Adapted from Ref.\cite{anissimova07}.} \label{fig:universal}
\end{figure*}

The two-parameter scaling theory \cite{punnoose01,punnoose05} of quantum diffusion in an interacting disordered system is based on the scaling hypothesis that both the resistivity and the electron-electron scattering amplitudes, $\gamma_2$, become scale (temperature) dependent. Essentially, this is the theory of Anderson localization in the presence of electron-electron interactions. The renormalization-group (RG) equations describing the evolution of the resistance and the scattering amplitude in 2D have the form~\cite{punnoose01}

\begin{eqnarray}
\frac{d\ln\rho^\ast}{d\xi}&=&\rho^\ast\;\left[n_v+1-(4n_v^2-1)\left(\frac{1+\gamma_2}{\gamma_2}\ln(1+\gamma_2)-1\right) \right],
\label{eqn:onelooprho}\\
\frac{d\gamma_2}{d\xi}&=&\rho^\ast\;\frac{(1+\gamma_2)^2}{2},
\label{eqn:oneloopgamma2}
\end{eqnarray}
where $\xi=-\ln(T\tau/\hbar)$, $\tau$ is the elastic scattering time, $\rho^\ast=(e^2/\pi h)\rho$, and $n_v$ is the number of degenerate valleys in the spectrum. The resistance, $\rho(T)$, is a non-monotonic function of temperature, reaching a maximum value $\rho_{max}$ at some temperature $T_{max}$ with a metallic temperature dependence ($d\rho/dT>0$) for $T<T_{max}$. Furthermore, $\rho(T)$ can be written as 
\begin{equation}
\rho = \rho_{max}\;F\left(\rho_{max}\;\text{ln}(T_{max}/T)\right),
\end{equation}
where $F$ is a \textit{universal} function shown by the solid curve in Fig. \ref{fig:universal}(a). The strength of spin-dependent interactions, $\gamma_2$, is also a universal function of $\text{ln}(T_{max}/T)$, shown by the solid line in Fig. \ref{fig:universal}(b).

This theory can account for the large changes in resistivity observed experimentally, and provides quantitative functions that can be directly compared with the experimental data. Such a comparison was made by Anissimova  \textit{et al.},\cite{anissimova07} who deduced the interaction amplitude from the magnetoresistance; the results are presented in Figs.~\ref{fig:universal}(a) and (b). The agreement between theory and experiment is especially striking given that the theory has no adjustable parameters. Systematic deviations from the universal curves occur at lower densities as higher order corrections in $\rho$ become important. Furthermore, the resistivity reaches a maximum at $\gamma_2\approx0.45$, in excellent agreement with theory \cite{punnoose01} for $n_v=2$.

\begin{figure*}
\centering \includegraphics[width=0.45\textwidth]{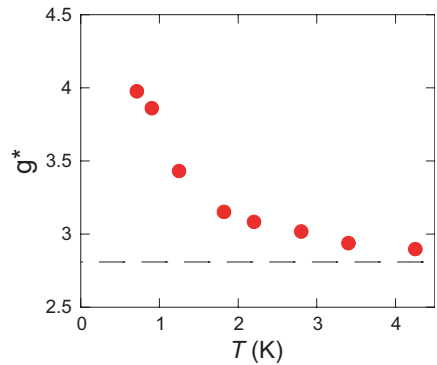}\\
\caption{g-factor \textit{vs.} temperature in the diffusive regime. The dashed line shows the value of the g-factor obtained in the ballistic regime. Adapted from Ref.\cite{krav_unp}.} \label{fig:g_factor}
\end{figure*}

The data for $\gamma_{2}$ enables one to calculate the renormalized Land\'{e} g-factor $g^{\ast }=2(1+\gamma _{2})$.  As shown in Fig.\ref{fig:g_factor}, at $n_{s}=9.87\times 10^{10}$~cm$^{-2}$, the Land\'{e} g-factor increases from $g^{\ast }\approx2.9$ at the highest temperature to $g^{\ast }\approx 4$ at the lowest. Therefore, the g-factor becomes temperature-dependent in the diffusive regime and increases with decreasing temperature, in agreement with the predictions of Punnoose and Finkelshtein\cite{punnoose01,punnoose05}. Note that at higher temperatures it is close to the temperature-independent ``ballistic'' value of about 2.8.

However, it should be noted that a theory based on electron interference effects can predict substantial changes of resistivity only in the near vicinity of the critical point, where $\rho\sim h/e^2$.  At much lower resistivities $\rho\ll h/e^2$, only small logarithmic corrections are possible, while the experiments demonstrate very large changes in resistivity even deep in the metallic region. Therefore, although the theory of Punnoose and Finkelstein \cite{punnoose01,punnoose05} quantitatively describes experimental data in the close vicinity of the transition, it cannot explain the large effects observed far from the transition. It should be further noted that most of the experiments in GaAs- and n-SiGe-based devices are performed in the ballistic regime where this theory is irrelevant.  Given the conspicuous similarity of the results obtained on different 2D systems, whether diffusive or ballistic, it seems clear that a unified theory applicable in both regimes is needed to account for the transport data.

Such a theory, based on the Pomeranchuk effect, was proposed by Spivak and Kivelson 
\cite{spivak03,spivak04,spivak06}. In essence, this theory proposes that when the interaction energy is much higher than the Fermi energy, the short-range interactions are of the Wigner-crystalline type. As the temperature or magnetic field is increased, the surface fraction occupied by the Wigner crystallites grows, explaining the metallic temperature dependence of the resistance and the giant positive magnetoresistance observed at low temperatures. Moreover, this theory predicts that the metallic temperature dependence of the resistance is quenched in magnetic fields strong enough to completely polarize the electrons' spins. The Pomeranchuk effect provides a qualitative explanation of all the major 
experimental observations; however, a quantitative theory is not currently available.

\subsection{Spin susceptibility and the effective mass enhancement}

There are several mechanism that could lead to the strong enhancement of the effective mass at low carrier densities (high $r_s$). Within Fermi liquid theory, the enhancement of $g$ and $m$ is due to spin exchange effects. Extension of the Fermi liquid model to relatively large $r_s$ \cite{iwamoto91,kwon94,chen99} is problematic, the main outcome being that the renormalization of $g$ is large compared to that of $m$. In the limiting case of high $r_s$, one may expect a divergence of the $g$ factor that corresponds to the Stoner instability. These predictions are in obvious contradiction to the experimental data. The divergence of the effective mass and spin susceptibility follow also from the Gutzwiller variational approach \cite{brinkman70} (see also Ref.\cite{dolgopolov02}). Recent theoretical developments include the following. Using a renormalization group analysis for multi-valley 2D systems, it has been found that the spin susceptibility increases dramatically as the density is decreased toward the critical density for the metal-insulator transition, while the $g$ factor remains nearly intact \cite{punnoose05}. However, this prediction is made for the diffusive regime, $T\tau/\hbar\ll1$, while the spin susceptibility enhancement has been observed well into the ballistic regime, $T\tau/\hbar\gg1$. In the Fermi-liquid-based model of Khodel \textit{et al}~\cite{khodel05}, a flattening at the Fermi energy in the spectrum that leads to a diverging effective mass has been predicted, in qualitative agreement with experiment, but a detailed microscopic theory is needed before conclusions can be drawn. The strong increase of the effective mass has also been obtained (in the absence of disorder) by solving an extended Hubbard model using dynamical mean-field theory \cite{pankov08,camjayi08}. This is consistent with experiment, especially taking into account that the relative mass enhancement has been experimentally found to be independent of the level of the disorder \cite{shashkin07}.

\section{Summary}
Although the behavior in the close vicinity of the transition is quantitatively described by the renormalization-group theory of Punnoose and Finkelstein \cite{punnoose05} without any fitting parameters, there is currently no microscopic theory that can explain the whole range of observed phenomena. The origin of the large changes in the resistance deep in the metallic phase remains unclear, with suggested explanations ranging from temperature-dependent screening \cite{dolgopolov00,dassarma99,dassarma00,dassarma03,dassarma04} to an analog of the Pomeranchuk effect \cite{spivak03,spivak04,spivak06}. From an empirical perspective, numerous experiments on various strongly correlated 2D electron and hole systems strongly suggest the existence of a metal-insulator transition and a metallic phase in two-dimensions, despite the persistent view on the part of many that such a transition is impossible in 2D.


\begin{thebibliography}{85}
\providecommand{\natexlab}[1]{#1}
\providecommand{\url}[1]{\texttt{#1}}
\expandafter\ifx\csname urlstyle\endcsname\relax
  \providecommand{\doi}[1]{doi: #1}\else
  \providecommand{\doi}{doi: \begingroup \urlstyle{rm}\Url}\fi

\bibitem{abrahams79}
E.~Abrahams, P.~W. Anderson, D.~C. Licciardello, and T.~V. Ramakrishnan,
  Scaling theory of localization: Absence of quantum diffusion in two
  dimensions, \emph{Phys.\ Rev.\ Lett.} {\bf 42}, \penalty0 673--676,  (1979).

\bibitem{finkelstein83}
A.~M. Finkel'shtein, Influence of {C}oulomb interaction on the properties of
  disordered metals, \emph{Sov.\ Phys.\ - JETP}. {\bf 57}, \penalty0 97--108,
  (1983).

\bibitem{finkelstein84b}
A.~M. Finkel'shtein, Weak localization and {C}oulomb interaction in
  disordered-systems, \emph{Z.\ Phys.\ B}. {\bf 56}, \penalty0 189--196,
  (1984).

\bibitem{castellani84}
C.~Castellani, C.~{Di~Castro}, P.~A. Lee, and M.~Ma, Interaction-driven
  metal-insulator transitions in disordered fermion systems, \emph{Phys.\ Rev.\
  B}. {\bf 30}, \penalty0 527--543,  (1984).

\bibitem{zavaritskaya87}
T.~N. Zavaritskaya and E.~I. Zavaritskaya, Metal-insulator transition in
  inversion channels of silicon {MOS} structures, \emph{JETP Lett.} {\bf 45},
  \penalty0 609--613,  (1987).

\bibitem{kravchenko94}
S.~V. Kravchenko, G.~V. Kravchenko, J.~E. Furneaux, V.~M. Pudalov, and
  M.~D'Iorio, Possible metal-insulator-transition at {$B=0$} in 2 dimensions,
  \emph{Phys.\ Rev.\ B}. {\bf 50}, \penalty0 8039--8042,  (1994).

\bibitem{kravchenko95a}
S.~V. Kravchenko, W.~E. Mason, G.~E. Bowker, J.~E. Furneaux, V.~M. Pudalov, and
  M.~D'Iorio, Scaling of an anomalous metal-insulator-transition in a
  2-dimensional system in silicon at {$B=0$}, \emph{Phys.\ Rev. B}. {\bf 51},
  \penalty0 7038--7045,  (1995).

\bibitem{kravchenko96}
S.~V. Kravchenko, D.~Simonian, M.~P. Sarachik, W.~Mason, and J.~E. Furneaux,
  Electric field scaling at a {$B=0$} metal-insulator transition in two
  dimensions, \emph{Phys.\ Rev.\ Lett.} {\bf 77}, \penalty0 4938--4941,
  (1996).

\bibitem{popovic97}
D.~Popovi\'c, A.~B. Fowler, and S.~Washburn, Metal-insulator transition in two
  dimensions: Effects of disorder and magnetic field, \emph{Phys.\ Rev.\ Lett.}
  {\bf 79}, \penalty0 1543--1546,  (1997).

\bibitem{mcfarland09}
R.~N. McFarland, T.~M. Kott, L.~Sun, K.~Eng, and B.~E. Kane,
  Temperature-dependent transport in a sixfold degenerate two-dimensional
  electron system on a {H-S}i(111) surface, \emph{Phys.\ Rev.\ B}. {\bf 80},
  \penalty0 161310(R),  (2009).

\bibitem{coleridge97}
P.~T. Coleridge, R.~L. Williams, Y.~Feng, and P.~Zawadzki, Metal-insulator
  transition at {$B=0$} in $p$-type {SiGe}, \emph{Phys.\ Rev.\ B}. {\bf 56},
  \penalty0 R12764--R12767,  (1997).

\bibitem{hanein98a}
Y.~Hanein, U.~Meirav, D.~Shahar, C.~C. Li, D.~C. Tsui, and H.~Shtrikman, The
  metalliclike conductivity of a two-dimensional hole system, \emph{Phys.\
  Rev.\ Lett.} {\bf 80}, \penalty0 1288--1291,  (1998).

\bibitem{hanein98b}
Y.~Hanein, D.~Shahar, J.~Yoon, C.~C. Li, D.~C. Tsui, and H.~Shtrikman,
  Observation of the metal-insulator transition in two-dimensional n-type
  {GaAs}, \emph{Phys.\ Rev.\ B}. {\bf 58}, \penalty0 R13338--R13340,  (1998).

\bibitem{papadakis98}
S.~J. Papadakis and M.~Shayegan, Apparent metallic behavior at {$B=0$} of a
  two-dimensional electron system in {AlAs}, \emph{Phys.\ Rev.\ B}. {\bf 57},
  \penalty0 R15068--R15071,  (1998).

\bibitem{okamoto04}
T.~Okamoto, M.~Ooya, K.~Hosoya, and S.~Kawaji, Spin polarization and metallic
  behavior in a silicon two-dimensional electron system, \emph{Phys.\ Rev.\ B}.
  {\bf 69}, \penalty0 041202(R),  (2004).

\bibitem{lai05}
K.~Lai, W.~Pan, D.~C. Tsui, S.~A. Lyon, M.~Muhlberger, and F.~Sch{\"a}ffler,
  Two-dimensional metal-insulator transition and in-plane magnetoresistance in
  a high-mobility strained {S}i quantum well, \emph{Phys.\ Rev.\ B}. {\bf 72},
  \penalty0 081313(R),  (2005).

\bibitem{dolgopolov92}
V.~T. Dolgopolov, G.~V. Kravchenko, A.~A. Shashkin, and S.~V. Kravchenko,
  Properties of electron insulating phase in {Si} inversion-layers at
  low-temperatures, \emph{JETP Lett.} {\bf 55}, \penalty0 733--737,  (1992).

\bibitem{simonian97b}
D.~Simonian, S.~V. Kravchenko, M.~P. Sarachik, and V.~M. Pudalov, Magnetic
  field suppression of the conducting phase in two dimensions, \emph{Phys.\
  Rev.\ Lett.} {\bf 79}, \penalty0 2304--2307,  (1997).

\bibitem{pudalov97}
V.~M. Pudalov, G.~Brunthaler, A.~Prinz, and G.~Bauer, Instability of the
  two-dimensional metallic phase to a parallel magnetic field, \emph{JETP
  Lett.} {\bf 65}, \penalty0 932--937,  (1997).

\bibitem{simmons98}
M.~Y. Simmons, A.~R. Hamilton, M.~Pepper, E.~H. Linfield, P.~D. Rose, D.~A.
  Ritchie, A.~K. Savchenko, and T.~G. Griffiths, Metal-insulator transition at
  {$B=0$} in a dilute two dimensional {GaAs-AlGaAs} hole gas, \emph{Phys.\
  Rev.\ Lett.} {\bf 80}, \penalty0 1292--1295,  (1998).

\bibitem{okamoto99}
T.~Okamoto, K.~Hosoya, S.~Kawaji, and A.~Yagi, Spin degree of freedom in a
  two-dimensional electron liquid, \emph{Phys.\ Rev.\ Lett.} {\bf 82},
  \penalty0 3875--3878,  (1999).

\bibitem{kravchenko00b}
S.~V. Kravchenko, A.~A. Shashkin, D.~A. Bloore, and T.~M. Klapwijk,
  Shubnikov-de {H}aas oscillations near the metal-insulator transition in a
  two-dimensional electron system in silicon, \emph{Solid State Commun.} {\bf
  116}, \penalty0 495--499,  (2000).

\bibitem{shashkin01a}
A.~A. Shashkin, S.~V. Kravchenko, V.~T. Dolgopolov, and T.~M. Klapwijk,
  Indication of the ferromagnetic instability in a dilute two-dimensional
  electron system, \emph{Phys.\ Rev.\ Lett.} {\bf 87}, \penalty0 086801,
  (2001).

\bibitem{vitkalov01b}
S.~A. Vitkalov, H.~Zheng, K.~M. Mertes, M.~P. Sarachik, and T.~M. Klapwijk,
  Scaling of the magnetoconductivity of silicon {MOSFETs}: Evidence for a
  quantum phase transition in two dimensions, \emph{Phys.\ Rev.\ Lett.} {\bf
  87}, \penalty0 086401,  (2001).

\bibitem{shashkin02}
A.~A. Shashkin, S.~V. Kravchenko, V.~T. Dolgopolov, and T.~M. Klapwijk, Sharp
  increase of the effective mass near the critical density in a metallic
  two-dimensional electron system, \emph{Phys.\ Rev.\ B}. {\bf 66}, \penalty0
  073303,  (2002).

\bibitem{vitkalov02}
S.~A. Vitkalov, M.~P. Sarachik, and T.~M. Klapwijk, Spin polarization of
  strongly interacting two-dimensional electrons: The role of disorder,
  \emph{Phys.\ Rev.\ B}. {\bf 65}, \penalty0 201106(R),  (2002).

\bibitem{pudalov02b}
V.~M. Pudalov, M.~E. Gershenson, H.~Kojima, N.~Butch, E.~M. Dizhur,
  G.~Brunthaler, A.~Prinz, and G.~Bauer, Low-density spin susceptibility and
  effective mass of mobile electrons in {Si} inversion layers, \emph{Phys.\
  Rev.\ Lett.} {\bf 88}, \penalty0 196404,  (2002).

\bibitem{zhu03}
J.~Zhu, H.~L. Stormer, L.~N. Pfeiffer, K.~W. Baldwin, and K.~W. West, Spin
  susceptibility of an ultra-low-density two-dimensional electron system,
  \emph{Phys.\ Rev.\ Lett.} {\bf 90}, \penalty0 056805,  (2003).

\bibitem{shashkin03}
A.~A. Shashkin, M.~Rahimi, S.~Anissimova, S.~V. Kravchenko, V.~T. Dolgopolov,
  and T.~M. Klapwijk, Spin-independent origin of the strongly enhanced
  effective mass in a dilute 2{D} electron system, \emph{Phys.\ Rev.\ Lett.}
  {\bf 91}, \penalty0 046403,  (2003).

\bibitem{anissimova06}
S.~Anissimova, A.~Venkatesan, A.~A. Shashkin, M.~R. Sakr, S.~V. Kravchenko, and
  T.~M. Klapwijk, Magnetization of a strongly interacting two-dimensional
  electron system in perpendicular magnetic fields, \emph{Phys.\ Rev.\ Lett.}
  {\bf 96}, \penalty0 046409,  (2006).

\bibitem{stoner46}
E.~C. Stoner, Ferromagnetism, \emph{Rep.\ Prog.\ Phys.} {\bf 11}, \penalty0
  43--112,  (1946).

\bibitem{punnoose05}
A.~Punnoose and A.~M. Finkel'stein, Metal-insulator transition in disordered
  two-dimensional electron systems, \emph{Science}. {\bf 310}, \penalty0
  289--291,  (2005).

\bibitem{senz99}
V.~Senz, U.~D{\"o}tsch, U.~Gennser, T.~Ihn, T.~Heinzel, K.~Ensslin,
  R.~Hartmann, and D.~Gr{\"u}tzmacher, Metal-insulator transition in a
  2-dimensional system with an easy spin axis, \emph{Ann.\ Phys.\ (Leipzig)}.
  {\bf 8}, \penalty0 237--240,  (1999).

\bibitem{yoon99}
J.~Yoon, C.~C. Li, D.~Shahar, D.~C. Tsui, and M.~Shayegan, Wigner
  crystallization and metal-insulator transition of two-dimensional holes in
  {GaAs} at {$B=0$}, \emph{Phys.\ Rev.\ Lett.} {\bf 82}, \penalty0 1744--1747,
  (1999).

\bibitem{mills99}
A.~P. Mills, Jr., A.~P. Ramirez, L.~N. Pfeiffer, and K.~W. West, Nonmonotonic
  temperature-dependent resistance in low density 2{D} hole gases, \emph{Phys.\
  Rev.\ Lett.} {\bf 83}, \penalty0 2805--2808,  (1999).

\bibitem{lilly03}
M.~P. Lilly, J.~L. Reno, J.~A. Simmons, I.~B. Spielman, J.~P. Eisenstein, L.~N.
  Pfeiffer, K.~W. West, E.~H. Hwang, and S.~{Das Sarma}, Resistivity of dilute
  2{D} electrons in an undoped {GaAs} heterostructure, \emph{Phys.\ Rev.\
  Lett.} {\bf 90}, \penalty0 056806,  (2003).

\bibitem{gao06}
X.~P.~A. Gao, G.~S. Boebinger, A.~P. Mills, Jr., A.~P. Ramirez, L.~N. Pfeiffer,
  and K.~W. West, Spin-polarization-induced tenfold magnetoresistivity of
  highly metallic two-dimensional holes in a narrow {GaAs} quantum well,
  \emph{Phys.\ Rev.\ B}. {\bf 73}, \penalty0 241315(R),  (2006).

\bibitem{lee85}
P.~A. Lee and T.~V. Ramakrishnan, Disordered electronic systems, \emph{Rev.\
  Mod.\ Phys.} {\bf 57}, \penalty0 287--337,  (1985).

\bibitem{kravchenko98}
S.~V. Kravchenko, D.~Simonian, M.~P. Sarachik, A.~D. Kent, and V.~M. Pudalov,
  Effect of tilted magnetic field on the anomalous {H}=0 conducting phase in
  high-mobility {Si MOSFETs}, \emph{Phys.\ Rev.\ B}. {\bf 58}, \penalty0
  3553--3556,  (1998).

\bibitem{pudalov02a}
V.~M. Pudalov, G.~Brunthaler, A.~Prinz, and G.~Bauer, Weak anisotropy and
  disorder dependence of the in-plane magnetoresistance in high-mobility (100)
  {S}i-inversion layers, \emph{Phys.\ Rev.\ Lett.} {\bf 88}, \penalty0 076401,
  (2002).

\bibitem{vitkalov00}
S.~A. Vitkalov, H.~Zheng, K.~M. Mertes, M.~P. Sarachik, and T.~M. Klapwijk,
  Small-angle {S}hubnikov-de~{H}aas measurements in a 2{D} electron system: The
  effect of a strong in-plane magnetic field, \emph{Phys.\ Rev.\ Lett.} {\bf
  85}, \penalty0 2164--2167,  (2000).

\bibitem{vitkalov01a}
S.~A. Vitkalov, M.~P. Sarachik, and T.~M. Klapwijk, Spin polarization of
  two-dimensional electrons determined from {S}hubnikov-de~{H}aas oscillations
  as a function of angle, \emph{Phys.\ Rev.\ B}. {\bf 64}, \penalty0 073101,
  (2001).

\bibitem{tutuc01}
E.~Tutuc, E.~P. {De~Poortere}, S.~J. Papadakis, and M.~Shayegan, In-plane
  magnetic field-induced spin polarization and transition to insulating
  behavior in two-dimensional hole systems, \emph{Phys.\ Rev.\ Lett.} {\bf 86},
  \penalty0 2858--2861,  (2001).

\bibitem{papadakis00}
S.~J. Papadakis, E.~P. {De~Poortere}, M.~Shayegan, and R.~Winkler, Anisotropic
  magnetoresistance of two-dimensional holes in {GaAs}, \emph{Phys.\ Rev.\
  Lett.} {\bf 84}, \penalty0 5592--5595,  (2000).

\bibitem{mertes01}
K.~M. Mertes, H.~Zheng, S.~A. Vitkalov, M.~P. Sarachik, and T.~M. Klapwijk,
  Temperature dependence of the resistivity of a dilute two-dimensional
  electron system in high parallel magnetic field, \emph{Phys.\ Rev.\ B}. {\bf
  63}, \penalty0 041101(R),  (2001).

\bibitem{shashkin01b}
A.~A. Shashkin, S.~V. Kravchenko, and T.~M. Klapwijk, Metal-insulator
  transition in 2{D}: equivalence of two approaches for determining the
  critical point, \emph{Phys.\ Rev.\ Lett.} {\bf 87}, \penalty0 266402,
  (2001).

\bibitem{gao02}
X.~P.~A. Gao, A.~P. Mills, Jr., A.~P. Ramirez, L.~N. Pfeiffer, and K.~W. West,
  Weak-localization-like temperature-dependent conductivity of a dilute
  two-dimensional hole gas in a parallel magnetic field, \emph{Phys.\ Rev.\
  Lett.} {\bf 88}, \penalty0 166803,  (2002).

\bibitem{tsui05}
Y.~Tsui, S.~A. Vitkalov, M.~P. Sarachik, and T.~M. Klapwijk, Conductivity of
  silicon inversion layers: Comparison with and without an in-plane magnetic
  field, \emph{Phys.\ Rev.\ B}. {\bf 71}, \penalty0 13308,  (2005).

\bibitem{landau57}
L.~D. Landau, The theory of a {F}ermi liquid, \emph{Sov.\ Phys.\ - JETP}. {\bf
  3}, \penalty0 920--925,  (1957).

\bibitem{fang68}
F.~F. Fang and P.~J. Stiles, Effects of a tilted magnetic field on a
  two-dimensional electron gas, \emph{Phys.\ Rev.} {\bf 174}, \penalty0
  823--828,  (1968).

\bibitem{smith72}
J.~L. Smith and P.~J. Stiles, Electron-electron interactions continuously
  variable in the range $2.1>r_s>0.9$, \emph{Phys.\ Rev.\ Lett.} {\bf 29},
  \penalty0 102--104,  (1972).

\bibitem{finkelstein84a}
A.~M. Finkel'shtein, Spin fluctuations in disordered systems near the
  metal-insulator transition, \emph{JETP Lett.} {\bf 40}, \penalty0 796--799,
  (1984).

\bibitem{prus02}
O.~Prus, Y.~Yaish, M.~Reznikov, U.~Sivan, and V.~M. Pudalov, Thermodynamic spin
  magnetization of strongly correlated two-dimensional electrons in a silicon
  inversion layer, \emph{Phys.\ Rev.\ B}. {\bf 67}, \penalty0 205407,  (2003).

\bibitem{shashkin06}
A.~A. Shashkin, S.~Anissimova, M.~R. Sakr, S.~V. Kravchenko, V.~T. Dolgopolov,
  and T.~M. Klapwijk, Pauli spin susceptibility of a strongly correlated 2{D}
  electron liquid, \emph{Phys.\ Rev.\ Lett.} {\bf 96}, \penalty0 036403,
  (2006).

\bibitem{vakili04}
K.~Vakili, Y.~P. Shkolnikov, E.~Tutuc, E.~P. {De~Poortere}, and M.~Shayegan,
  Spin susceptibility of two-dimensional electrons in narrow {AlAs} quantum
  wells, \emph{Phys.\ Rev.\ Lett.} {\bf 92}, \penalty0 226401,  (2004).

\bibitem{lu08}
T.~M. Lu, L.~Sun, D.~C. Tsui, S.~Lyon, W.~Pan, M.~Muhlberger, F.~Sch{\"a}ffler,
  J.~Liu, and Y.~H. Xie, In-plane field magnetoresistivity of {S}i
  two-dimensional electron gas in {Si/SiGe} quantum wells at 20 m{K},
  \emph{Phys.\ Rev.\ B}. {\bf 78}, \penalty0 233309,  (2008).

\bibitem{shashkin01}
A.~A. Shashkin, S.~V. Kravchenko, and T.~M. Klapwijk, Metal-insulator
  transition in {2D}: equivalence of two approaches for determining the
  critical point, \emph{Phys.\ Rev.\ Lett.} {\bf 87}, \penalty0 266402,
  (2001).

\bibitem{ando82}
T.~Ando, A.~B. Fowler, and F.~Stern, Electronic-properties of two-dimensional
  systems, \emph{Rev.\ Mod.\ Phys.} {\bf 54}, \penalty0 437--672,  (1982).

\bibitem{iwamoto91}
N.~Iwamoto, Static local-field corrections of two-dimensional electron liquids,
  \emph{Phys.\ Rev.\ B}. {\bf 43}, \penalty0 2174--2182,  (1991).

\bibitem{kwon94}
Y.~Kwon, D.~M. Ceperley, and R.~M. Martin, Quantum {M}onte {C}arlo calculation
  of the {F}ermi-liquid parameters in the two-dimensional electron gas,
  \emph{Phys.\ Rev.\ B}. {\bf 50}, \penalty0 1684--1694,  (1994).

\bibitem{chen99}
G.-H. Chen and M.~E. Raikh, Exchange-induced enhancement of spin-orbit coupling
  in two-dimensional electronic systems, \emph{Phys.\ Rev.\ B}. {\bf 60},
  \penalty0 4826--4833,  (1999).

\bibitem{shashkin07}
A.~A. Shashkin, A.~A. Kapustin, E.~V. Deviatov, V.~T. Dolgopolov, and Z.~D.
  Kvon, Strongly enhanced effective mass in dilute two-dimensional electron
  systems: System-independent origin, \emph{Phys.\ Rev.\ B}. {\bf 76},
  \penalty0 241302(R),  (2007).

\bibitem{phillips98}
P.~Phillips, Y.~Wan, I.~Martin, S.~Knysh, and D.~Dalidovich, Superconductivity
  in a two-dimensional electron gas, \emph{Nature (London)}. {\bf 395},
  \penalty0 253--257,  (1998).

\bibitem{chakravarty99}
S.~Chakravarty, S.~Kivelson, C.~Nayak, and K.~Voelker, Wigner glass,
  spin-liquids, and the metal-insulator transition, \emph{Phil.\ Mag.\ B}. {\bf
  79}, \penalty0 859--868,  (1999).

\bibitem{spivak03}
B.~Spivak, Phase separation in the two-dimensional electron liquid in
  {MOSFETs}, \emph{Phys.\ Rev.\ B}. {\bf 67}, \penalty0 125205,  (2003).

\bibitem{spivak04}
B.~Spivak and S.~A. Kivelson, Intermediate phases of the two dimensional
  electron fluid between the {F}ermi liquid and the {W}igner crystal,
  \emph{Phys.\ Rev.\ B}. {\bf 70}, \penalty0 155114,  (2004).

\bibitem{spivak06}
B.~Spivak and S.~A. Kivelson, Transport in two dimensional electronic
  micro-emulsions, \emph{Ann.\ Phys.} {\bf 321}, \penalty0 2071--2115,  (2006).

\bibitem{meir99}
Y.~Meir, Percolation-type description of the metal-insulator transition in two
  dimensions, \emph{Phys.\ Rev.\ Lett.} {\bf 83}, \penalty0 3506â€“3509,
  (1999).

\bibitem{chakravarty98}
S.~Chakravarty, L.~Yin, and E.~Abrahams, Interactions and scaling in a
  disordered two-dimensional metal, \emph{Phys.\ Rev.\ B}. {\bf 58}, \penalty0
  R559--R562,  (1998).

\bibitem{stern80}
F.~Stern, Calculated temperature dependence of mobility in silicon inversion
  layers, \emph{Phys.\ Rev.\ Lett.} {\bf 44}, \penalty0 1469--1472,  (1980).

\bibitem{gold86}
A.~Gold and V.~T. Dolgopolov, Temperature dependence of the conductivity for
  the two-dimensional electron gas: Analytical results for low temperatures,
  \emph{Phys.\ Rev.\ B}. {\bf 33}, \penalty0 1076--1084,  (1986).

\bibitem{dassarma99}
S.~{Das~Sarma} and E.~H. Hwang, Charged impurity scattering limited low
  temperature resistivity of low density silicon inversion layers, \emph{Phys.\
  Rev.\ Lett.} {\bf 83}, \penalty0 164--167,  (1999).

\bibitem{dolgopolov00}
V.~T. Dolgopolov and A.~Gold, Magnetoresistance of a two-dimensional electron
  gas in a parallel magnetic field, \emph{JETP Lett.} {\bf 71}, \penalty0
  27--30,  (2000).

\bibitem{dassarma00}
S.~{Das~Sarma} and E.~H. Hwang, Parallel magnetic field induced giant
  magnetoresistance in low density quasi-two-dimensional layers, \emph{Phys.\
  Rev.\ B}. {\bf 61}, \penalty0 R7838--R7841,  (2000).

\bibitem{dassarma03}
S.~{Das~Sarma} and E.~H. Hwang, Low density finite temperature apparent
  insulating phase in 2{D} semiconductor systems, \emph{Phys.\ Rev.\ B}. {\bf
  68}, \penalty0 195315,  (2003).

\bibitem{dassarma04}
S.~{Das~Sarma} and E.~H. Hwang, Metallicity and its low temperature behavior in
  dilute 2{D} carrier systems, \emph{Phys.\ Rev.\ B}. {\bf 69}, \penalty0
  195305,  (2004).

\bibitem{zala01}
G.~Zala, B.~N. Narozhny, and I.~L. Aleiner, Interaction corrections at
  intermediate temperatures: Longitudinal conductivity and kinetic equation,
  \emph{Phys.\ Rev.\ B}. {\bf 64}, \penalty0 214204,  (2001).

\bibitem{punnoose01}
A.~Punnoose and A.~M. Finkelstein, Dilute electron gas near the metal-insulator
  transition: Role of valleys in silicon inversion layers, \emph{Phys.\ Rev.\
  Lett.} {\bf 88}, \penalty0 016802,  (2001).

\bibitem{anissimova07}
S.~Anissimova, S.~V. Kravchenko, A.~Punnoose, A.~M. Finkel'stein, and T.~M.
  Klapwijk, Flow diagram of the metal-insulator transition in two dimensions,
  \emph{Nature Phys.} {\bf 3}, \penalty0 707--710,  (2007).

\bibitem{krav_unp}
A.~Mokashi and S.~V. Kravchenko, \emph{unpublished}.  (2009).

\bibitem{brinkman70}
W.~F. Brinkman and T.~M. Rice, Application of {G}utzwiller's variational method
  to the metal-insulator transition, \emph{Phys.\ Rev.\ B}. {\bf 2}, \penalty0
  4302--4304,  (1970).

\bibitem{dolgopolov02}
V.~T. Dolgopolov, On effective electron mass of silicon field structures at low
  electron densities, \emph{JETP Lett.} {\bf 76}, \penalty0 377--379,  (2002).

\bibitem{khodel05}
V.~A. Khodel, J.~W. Clark, and M.~V. Zverev, Thermodynamic properties of
  {F}ermi systems with flat single-particle spectra, \emph{Europhys.\ Lett.}
  {\bf 72}, \penalty0 256--262,  (2005).

\bibitem{pankov08}
S.~Pankov and V.~Dobrosavljevi\'c, Self-doping instability of the
  {W}igner-{M}ott insulator, \emph{Phys.\ Rev.\ B}. {\bf 77}, \penalty0 085104,
   (2008).

\bibitem{camjayi08}
A.~Camjayi, K.~Haule, V.~Dobrosavljevi\'c, and G.~Kotliar, Coulomb correlations
  and the {W}igner|-{M}ott transition, \emph{Nature Phys.} {\bf 4}, \penalty0
  932--935,  (2008).

\end{thebibliography}

\end{document}